\documentclass[11pt,draftclsnofoot,onecolumn]{IEEEtran}

\usepackage[dvips]{graphicx}
\usepackage{mathrsfs}
\usepackage{amssymb}
\usepackage{latexsym}
\usepackage{multirow}
\usepackage[cmex10]{amsmath}
\usepackage{caption}
\usepackage[font=footnotesize]{subfig}

\begin{document}

\title{Medium Access Control Protocols\\With Memory}

\author{Jaeok Park and Mihaela van der Schaar\thanks{The authors are with Electrical Engineering Department, University of
California, Los Angeles (UCLA), 420 Westwood Plaza,
Los Angeles, CA 90095-1594, USA. e-mail:
\{jaeok, mihaela\}@ee.ucla.edu.}}

\maketitle

\begin{abstract}
Many existing medium access control (MAC) protocols utilize past
information (e.g., the results of transmission attempts)
to adjust the transmission parameters of users. This paper provides a
general framework to express and evaluate distributed MAC protocols utilizing
a finite length of memory for a given form of feedback information.
We define protocols with memory in the context of a slotted random
access network with saturated arrivals. We introduce two performance
metrics, throughput and average delay, and formulate the problem of
finding an optimal protocol. We first show that a TDMA outcome,
which is the best outcome in the considered scenario,
can be obtained after a transient period by a protocol with
$(N-1)$-slot memory, where $N$ is the total number of
users.
Next, we analyze the performance of protocols with 1-slot memory
using a Markov chain and numerical methods. Protocols with 1-slot memory
can achieve throughput arbitrarily close to 1 (i.e., 100\% channel utilization) at the expense of large average delay,
by correlating successful users in two consecutive slots.
Finally, we apply our framework to wireless local area
networks.
\end{abstract}


\section{Introduction}

In multiaccess communication systems, multiple
users share a communication channel and contend for access. Medium
access control (MAC) protocols are used to coordinate access and
resolve contention among users. We can categorize MAC protocols\footnote{%
Since we deal with MAC protocols exclusively in this paper,
we use the term ``protocol'' to represent ``MAC protocol'' hereafter.}
into two classes, centralized and distributed protocols, depending on the existence of
a central entity that coordinates the transmissions of users. Time division multiple access (TDMA)
is an example of centralized protocols,
where a scheduler assigns time slots to users. Centralized control
can achieve a high level of channel utilization by avoiding collisions,
but it requires large overhead for the communication of control messages.
Slotted Aloha and IEEE 802.11 distributed
coordination function (DCF) are examples of distributed protocols.
In slotted Aloha, users transmit new packets in the next time slot
while retransmitting backlogged packets with a fixed probability.
In DCF, carrier sense multiple access with
collision avoidance (CSMA/CA) and binary slotted exponential backoff (EB)
are used for users to determine their transmission times.
These distributed protocols can
be implemented without explicit control messages, but coordination is
limited in that collisions may occur or the channel may be unused when
some users have packets to send.

In this paper, we aim to improve the degree of coordination attainable
with distributed protocols by introducing \emph{memory} into the MAC
layer. Under a protocol with memory, a node dynamically adjusts its transmission
parameters based on the history of its local information.
The idea of utilizing histories at the MAC layer can be found in
various existing protocols. For example,
the slotted Aloha protocol \cite{rob} and its generalized version \cite{ma}
adjust the transmission probabilities of nodes depending on whether the current
packet is new or backlogged. The pseudo-Bayesian algorithm of \cite{riv}
utilizes channel feedback to update the estimated number of backlogged
packets in the system, based on which the transmission probability
is determined. The EB protocols in \cite{lee} use the results of transmission
attempts to adjust the contention window and current window sizes of nodes
or their transmission probabilities.

The above protocols, however, utilize available past information
in a limited way. We can consider the
above protocols as the current state of a node determining its transmission
parameters and the state transition occurring based on its local observations.
Although this structure makes implementation simple in that nodes
can simply keep track of their states in order to make transmission
decisions, there may be many possible paths that lead to the same
state, and important information may be lost by aggregating different histories into a single state.
For example, in slotted
Aloha, a node with a backlogged packet uses the same transmission
probability following a slot in which it waited and following a slot in which
it transmitted and collided. However, these two outcomes are observable
by the node, and a significant performance improvement may be achieved by using different transmission probabilities following
the two outcomes.
Another limitation of the above protocols is that they are designed assuming
a particular form of feedback  information. In case that more informative feedback is available,
utilizing the additional information may result in performance gains.
For example, the EB protocols in \cite{lee} prescribe that a node
should not update its parameters following a waiting slot. If a
node can sense the channel while waiting, utilizing the information
obtained from sensing may improve the performance of the EB protocols.

In order to overcome the limitations of the existing protocols utilizing memory,
we provide a systematic framework to express and evaluate protocols
with memory in the context of a slotted multiaccess system with saturated
arrivals where nodes make transmission decisions based on their transmission probabilities.
Our framework allows us to formally express a protocol utilizing memory of
any finite length and operating under any form of feedback information.
Also, we introduce two performance metrics, throughput and average delay,
based on which we can evaluate protocols with memory.
The two main results of this paper can be summarized as follows.
\begin{enumerate}
\item In the considered scenario with saturated arrivals, TDMA is the best
protocol in a sense that there is no other protocol that achieves a higher throughput
or a smaller average delay. A TDMA
outcome can be obtained after a transient period by a protocol with
$(N-1)$-slot or $N$-slot memory, where $N$ is the total number of nodes in the system.
\item A protocol with 1-slot memory can achieve throughput arbitrarily
close to 1 (i.e., 100\% channel utilization) at the expense of large average delay, by correlating successful
nodes in two consecutive slots (i.e., a node that has a successful transmission
in the current slot has a high probability of success in the next slot).
\end{enumerate}

The proposed protocols with memory can be related to splitting algorithms
\cite{bert} and reservation Aloha \cite{crowther}. In splitting
algorithms such as tree algorithms \cite{cape}, backlogged nodes are
divided into groups, one of which transmits in the next slot.
Protocols with memory use histories to split nodes into groups.
As nodes randomly access the channel based on transmission probabilities, histories
will evolve differently across nodes as time passes.
The probability of a successful transmission can be made high
by choosing transmission probabilities in a way that
the expected size of the transmitting group is approximately one
most of the time. In reservation Aloha, nodes maintain frames with
a certain number of slots, and a successful transmission serves as a reservation
for the same slot in the next frame. Reservation
Aloha can thus be expressed as a protocol with memory whose length is
equal to the number of slots in a frame, provided that all nodes
can learn successful transmissions in the system.
Protocols with memory are more flexible than reservation Aloha
in that protocols with memory can specify different
transmission probabilities in non-reserved slots, can make reservations
in a probabilistic way, and can be implemented with an arbitrary
form of feedback information.

The rest of this paper is organized as follows. In Section II, we
describe the considered slotted multiaccess model and define protocols
with memory. In Section III, we introduce performance metrics and
formulate the problem of finding an optimal protocol.
In Section IV, we show that a protocol with $(N-1)$-slot or $N$-slot memory can achieve
the performance of TDMA.
In Section V, we analyze the properties of protocols with 1-slot memory
using numerical methods. In Section VI, we show that protocols with
memory can be applied to the wireless local area
network (WLAN) environment and can achieve a performance improvement over DCF.
We conclude the paper in Section VII.

\section{System Model}

\subsection{Setup}

We consider a slotted multiaccess system as in
\cite{ma}. The system has total $N$ contending users, or transmitter nodes,
and the set of users is denoted by $\mathcal{N}
= \{ 1, \ldots, N \}$. We assume that the number of users is fixed
over time and known to users. Users share a communication channel
through which they transmit packets. Time is slotted, and users are
synchronized in their slot transmission times. We label slots by $t = 1,2,\ldots$. A user always has
a packet to transmit and can attempt to transmit one packet in each
slot. The set of actions available to a user in a slot is denoted by $A
\triangleq \{T,W\}$, where $T$ stands for ``transmit'' and $W$ for
``wait.'' We denote the action of user $i$ by $a_i \in A$ and an
action profile, or \emph{a transmission outcome}, by $\mathbf{a} \triangleq
(a_1,\ldots,a_N)$. The set of (transmission) outcomes is denoted by $\mathcal{A}
\triangleq A^N$. A transmission is successful if it is the
only transmission in the slot, and two or more transmissions in the
same slot result in a collision.

\subsection{Feedback Information}

After a user transmits a packet, it learns whether the packet is
successfully transmitted or not using
an acknowledgement (ACK) response. If the user receives an ACK from the
receiver node, it learns that its transmission was successful.
Otherwise, it concludes that its packet has collided. We assume that
there is no error in the transmission and the reception of ACK responses
so that a user always learns the correct results of its transmission attempts.
Formally, we represent ACK feedback to user $i$ by $ACK_i(\mathbf{a})$, which takes the value
$yes$ if $\mathbf{a} = \mathbf{a}^i$, where $\mathbf{a}^i$ is the outcome in which only user $i$
transmits, and $no$ otherwise.

At the end of each slot, users obtain channel feedback about the number of transmissions
in that slot. Let $k(\mathbf{a})$ be the number of transmissions
in outcome $\mathbf{a}$. The set of possible numbers
of transmissions in a slot is $\mathcal{K} \triangleq \{0,1, \ldots, N \}$.
Let $H_i$ be an information partition of $\mathcal{K}$ for user $i$ \cite{fudenberg}.
Then channel feedback to user $i$ in a slot with outcome $\mathbf{a}$ is given by the element of $H_i$
that contains $k(\mathbf{a})$, which we denote by $h_i(\mathbf{a})$.
The definition of information partitions requires that
$k(\mathbf{a}) \in h_i(\mathbf{a})$ for all $\mathbf{a} \in \mathcal{A}$. That is, channel
feedback never leads users to regard the actual number of transmissions as impossible.
However, when channel feedback has errors, it is possible for users
to obtain incorrect channel feedback such that $k(\mathbf{a}) \notin h_i(\mathbf{a})$. In the presence of
the hidden terminal problem (also called ``erasures'' in \cite{bing}),
users may interpret a success or a collision slot as an idle slot.
Under noise errors \cite{bing}, users may interpret an idle or a success
slot as a collision slot. It is also possible that users obtain channel feedback
in a probabilistic way, in which case $h_i(\mathbf{a})$ is determined
according to a probability distribution on $H_i$, $\triangle(H_i)(\mathbf{a})$.

All the aforementioned channel feedback settings can be incorporated in the formulation
of the protocol designer's problem developed in Section III. Although modeling
general settings can be of importance
in practice, we impose the following simplifying assumptions on channel feedback
for analytic convenience.
First, we assume that channel feedback is generated in a deterministic way. Then
channel feedback to user $i$ can be represented by a mapping $h_i$
from $\mathcal{A}$ to $H_i$, instead of $\triangle(H_i)$. Second, we
assume that $k(\mathbf{a}) \in h_i(\mathbf{a})$ for all $\mathbf{a} \in
\mathcal{A}$ so that channel feedback always contains the actual
number of transmissions. Finally, we assume that every user obtains
the same channel feedback, i.e., $h_1 = \cdots = h_N \triangleq h$,
which requires $H_1 = \cdots = H_N \triangleq H$.
Under these three assumptions, a channel feedback model is completely
described by an information partition $H$ of $\mathcal{K}$.

Consider a partition of $\mathcal{K}$, $\{\{0\},\{1\},e\}$, where
$e \triangleq \{2,\ldots,N\}$.
Each element of the partition represents a type of channel feedback, and users can potentially learn whether
there has been zero packet (idle), one packet (success), or more than one
packet (collision) transmitted in the slot. With an abuse of notation, we will use 0 and 1 to represent
channel feedback that corresponds to $\{0\}$ and $\{1\}$,
respectively \cite{bert}.
We say that the channel possesses ternary feedback, or $(0, 1, e)$ feedback, if all three
types of feedback are available. Binary feedback is also possible, and
given the three possible types of channel feedback, we can consider three kinds of binary feedback:
success/failure (S/F) feedback, which informs users whether there was a successful transmission (1)
or not ($0 \cup e$); collision/no collision (C/NC) feedback, which informs users whether there
was a collision ($e$) or not ($0 \cup 1$); and empty/not empty (E/NE) feedback, which informs users
whether the current slot was empty (0) or not ($1 \cup e$) \cite{bing}.
We can also consider no channel feedback, which does not give any channel information
to users and corresponds to information partition $\{\mathcal{K}\}$,
and $(N+1)$-ary feedback, which informs users of the exact number
of transmissions and corresponds to information partition $\{ \{0\},\{1\},\ldots,\{N\} \}$,
the finest partition of $\mathcal{K}$.
(See \cite{mahravari} for a similar list of channel feedback models in a multiple
reception scenario.)

The \emph{feedback information} of user $i$ consists of ACK feedback and channel
feedback, i.e., $(ACK_i(\mathbf{a}), h_i(\mathbf{a}))$.
We define a \emph{feedback technology} as a rule that
generates feedback information for each user depending on transmission outcomes.
The above three assumptions on channel feedback allow us to represent a feedback technology
by a mapping $\rho:\mathcal{S} \rightarrow 2^{\mathcal{K}}$, where
$\mathcal{S} \triangleq \cup_{\mathbf{a} \in \mathcal{A}} \{(a_i, k(\mathbf{a}))\}$,
and write the feedback information of
user $i$ more compactly as $\rho(a_i, k(\mathbf{a}))$.
Since $ACK_i(\mathbf{a}) = no$ if $a_i = W$, we can set $\rho(W, k(\mathbf{a}))
= h(\mathbf{a})$. Since user $i$ can distinguish the outcomes in 1
from those in $e$ using ACK feedback whenever it transmits, we can set $\rho(T, k(\mathbf{a}))
= 1$ if $ACK_i(\mathbf{a}) = yes$ and $h(\mathbf{a}) \cap \, e$
if $ACK_i(\mathbf{a}) = no$. Let $Z_{\rho} \triangleq \cup_{\mathbf{a} \in \mathcal{A}} \{\rho(a_i, k(\mathbf{a}))\}$,
which is independent of $i$ by the symmetry assumption.
$Z_{\rho}$ represents the set of feedback information that a user can obtain
with feedback technology $\rho$. The feedback information
of user $i$ is denoted by  $z_i \in Z_{\rho}$.
We use $\mathcal{R}$ to denote the set of all
feedback technologies,
which is equivalent to the set of all partitions of $\mathcal{K}$.

\subsection{Protocols With Memory}

A user decides whether to transmit or not in each slot using a transmission
probability, which lies in $[0,1]$. A protocol is a rule
based on which users determine their transmission probabilities.
We assume that control or coordination messages cannot be used
in the system. Then the
transmission action and the feedback information of user $i$ are
all the information that it obtains in a slot. The \emph{$M$-slot history}
for user $i$ in slot $t$ is given by
\begin{align*}
L_i^t = (a_i^{t-M}, z_i^{t-M}; \ldots; a_i^{t-1}, z_i^{t-1}),
\end{align*}
for $M = 1,2,\ldots$ and $t = 1,2,\ldots$.\footnote{In addition to actions and feedback,
a user knows the transmission probabilities it has used. We do not include
past transmission probabilities in histories because we focus on protocols that do not depend
on past transmission probabilities directly.} We set $(a_i^{t'},
z_i^{t'}) = (W, \rho(W,0))$ for $t' \leq 0$ as initialization. Let
$\mathcal{L}_{\rho}$ be the set of possible action-feedback pairs
under feedback technology $\rho$, i.e.,
$\mathcal{L}_{\rho} \triangleq \cup_{\mathbf{a} \in \mathcal{A}} \{(a_i,\rho(a_i, k(\mathbf{a})))\} \subset A \times Z_{\rho}$.
Then the set of $M$-slot histories is given by
$\mathcal{L}_{\rho}^{M}$. A stationary decision rule based on
$M$-slot histories is defined by a mapping
\begin{align*}
f: \mathcal{L}_{\rho}^{M} \rightarrow [0,1],
\end{align*}
where $f(L)$ represents the transmission probability for a user whose
$M$-slot history is $L \in \mathcal{L}_{\rho}^{M}$.

Since the set of $M$-slot histories $\mathcal{L}_{\rho}^{M}$ is affected
by feedback technology $\rho$, the set of stationary decision
rules based on $M$-slot histories depends
on the feedback technology of the system. In particular, as the
feedback technology is more informative in a sense that the corresponding
information partition is finer, users can distinguish more
outcomes, and thus more decision rules can be deployed. We use
$\mathcal{F}_{M,\rho}$ to denote the set of all stationary decision
rules based on $M$-slot histories under feedback technology $\rho$.
We define a \emph{protocol with $M$-slot memory} as a profile of
stationary decision rules based on $M$-slot histories given a
feedback technology $\rho$, i.e., $\mathbf{f}
\triangleq (f_1, \ldots, f_N) \in \mathcal{F}_{M,\rho}^N$. We say
that a protocol $\mathbf{f}$ is \emph{symmetric} if it prescribes
the same decision rule to every user, i.e., $f_1 = \cdots = f_N$.
We will sometimes use $f$ to
represent a symmetric protocol $\mathbf{f}$ with a common decision
rule $f$ when there is no confusion.

\subsection{Automaton Representations of Protocols With Memory}

A protocol with memory can be described by
a finite automaton, which consists of a finite set of states, an initial state, an action rule,
and a state transition rule \cite{mailath}. We can represent a symmetric
protocol $f \in \mathcal{F}_{M,\rho}$ as a finite automaton
by defining states as possible $M$-slot histories, the initial state as
the $M$-slot history obtained from $M$ idle slots, the action rule
as the decision rule $f$, and the state transition rule as to specify
the new state as the $M$-slot history updated based on the transmission action
and the feedback information in the current slot.

\begin{figure}%
\centering
\subfloat[][]{%
\label{fig:auto-a}%
\includegraphics[width=0.4\textwidth]{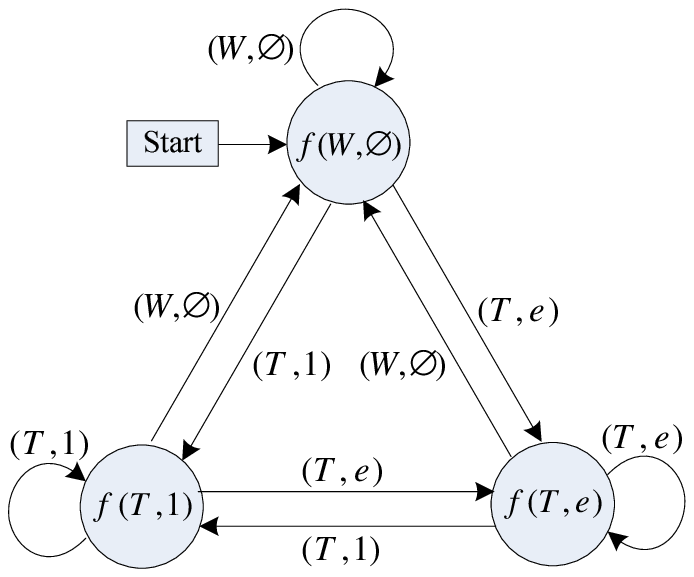}}%
\subfloat[][]{%
\label{fig:auto-b}%
\includegraphics[width=0.52\textwidth]{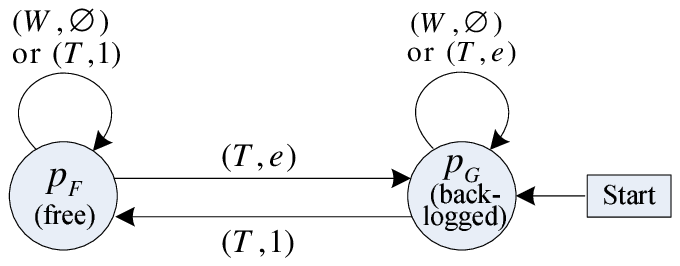}}%
\caption[Automaton representation.]{Automaton representations of protocols:
\protect\subref{fig:auto-a} a protocol with 1-slot memory, and
\protect\subref{fig:auto-b} a two-state protocol.
}%
\label{fig:automata}%
\end{figure}

Fig.~\ref{fig:automata}\subref{fig:auto-a} shows the automaton representation
of a protocol with memory in the simplest case
of 1-slot memory and no channel feedback (i.e., ACK feedback only).
With ACK feedback only, there are three possible
action-feedback pairs, or 1-slot histories: $(W,\emptyset)$, $(T,1)$, and $(T,e)$, where
$\emptyset$ corresponds to channel feedback $\mathcal{K}$.
Hence, there are three states in the automaton, corresponding to
the three possible action-feedback pairs, with the initial state
specified as $(W,\emptyset)$. The transmission probability
in each state $(a,z)$ is given by $f(a,z)$. The state in the next slot
is determined by the action-feedback pair in the current slot.

\cite{ma} proposes generalized slotted Aloha protocols, which
we call two-state protocols. Under a two-state
protocol, users choose their transmission probabilities
depending on whether their current packets are new (a free state) or
have collided before (a backlogged state). Thus, having ACK feedback
suffices to implement a two-state protocol. The automaton representation
of a two-state protocol is shown in Fig.~\ref{fig:automata}\subref{fig:auto-b}.
In general, a two-state protocol
cannot be expressed as a protocol with finite memory. Since the state
remains the same following a waiting slot, the transmission
probability for the current slot is determined by the result
of the most recent transmission attempt, which may
have occurred arbitrarily many slots ago. However,
\cite{ma} shows that under saturated arrivals two-state protocols achieve the maximum throughput when $p_F = 1$ and
$p_G \rightarrow 0$, where $p_F$ and $p_G$ are transmission
probabilities in the free and backlogged states, respectively. When a two-state
protocol specifies $p_F = 1$, which is also the case in the original slotted
Aloha protocol, the action-feedback pair $(W,\emptyset)$
cannot occur in the free state and thus we can find an equivalent protocol
with 1-slot memory such that $f(T,1) = 1$ and
$f(W,\emptyset) = f(T,e) = p_G$. In effect, a two-state
protocol with $p_F = 1$ aggregates the two different action-feedback
pairs, $(W,\emptyset)$ and $(T,e)$, into one state, the backlogged state,
and assigns the same transmission probability following these two action-feedback
pairs. On the contrary, a protocol with 1-slot
memory can fully utilize all the available information from the
previous slot in that three different transmission probabilities can be assigned
following the three action-feedback pairs.

As will be shown later in Section V.D., the limited utilization of past information by
two-state protocols results in performance limitations. In particular,
throughput, or the fraction of success slots, is bounded from above by $N/(2N-1)$ under two-state protocols (Theorem 2 of \cite{ma}) while
protocols with 1-slot memory can achieve throughput arbitrarily close to 1.
When memory longer than 1 slot or channel feedback is available,
there are more distinguishable histories for each user.
Two-state protocols aggregate different histories
into just two states whereas protocols with memory can assign as many
transmission probabilities as the number of possible histories given the length of memory and
the feedback technology.
Thus, the performance gap between
the two kinds of protocols will be larger when longer memory or more informative
feedback is available.

\section{Problem Formulation}

\subsection{Performance Metrics}

\subsubsection{Throughput}

We define the \emph{throughput} of a user as the fraction
of slots in which it has a successful transmission and \emph{total throughput} as
the fraction of slots in which there is a successful transmission
in the system, which is equal to the sum of the individual throughput of users.
When users follow a protocol with memory $\mathbf{f}$,
throughput can be computed using a Markov chain. Let $M$ be
the length of memory used by protocol $\mathbf{f}$ and $\rho$ be
the associated feedback technology. Then we can consider a Markov chain
whose state space is given by $\mathcal{A}^{M}$. We write an element of $\mathcal{A}^{M}$
as $\bar{\mathbf{a}} \triangleq (\mathbf{a}_1, \ldots,
\mathbf{a}_{M})$. Let $L_i(\bar{\mathbf{a}})$ be the $M$-slot
history for user $i$ when the outcomes in the recent $M$ slots are
$\bar{\mathbf{a}}$, i.e.,
\begin{align*}
L_i(\bar{\mathbf{a}}) = (a_{i,1}, \rho(a_{i,1},k(\mathbf{a}_1));
\ldots; a_{i,M}, \rho(a_{i,M},k(\mathbf{a}_{M}))),
\end{align*}
where $a_{i,m}$ is the $i$th element of $\mathbf{a}_{m}$. Given
$L_i(\bar{\mathbf{a}})$ as its $M$-slot history, user $i$ transmits
with probability $f_i(L_i(\bar{\mathbf{a}}))$. The transmission
probabilities of users yield a probability distribution on the
outcome in the current slot. The probability that the current
outcome is $\mathbf{a}'$ under $\mathbf{f}$ when the outcomes in the
recent $M$ slots are $\bar{\mathbf{a}}$ is given by
\begin{align} \label{eq:trans23}
P(\mathbf{a}'|\bar{\mathbf{a}};\mathbf{f}) = \prod_{i=1}^N \left[
\mathbf{1}_{\{\mathbf{a}:a_{i} = T\}}(\mathbf{a}')
f_i(L_i(\bar{\mathbf{a}})) + \mathbf{1}_{\{\mathbf{a}:a_{i} = W\}}(\mathbf{a}')
(1-f_i(L_i(\bar{\mathbf{a}}))) \right],
\end{align}
where $\mathbf{1}_{\mathcal{B}}:\mathcal{A} \rightarrow \{0,1\}$ is
an indicator function such that
$\mathbf{1}_{\mathcal{B}}(\mathbf{a}) = 1$ if $\mathbf{a} \in
\mathcal{B}$ and 0 otherwise. The transition probability from
$\bar{\mathbf{a}} \in \mathcal{A}^{M}$ to $\bar{\mathbf{a}}'
\triangleq (\mathbf{a}'_1, \ldots, \mathbf{a}'_{M}) \in
\mathcal{A}^{M}$ under protocol $\mathbf{f}$ is given by
\begin{eqnarray*}
Q(\bar{\mathbf{a}}'|\bar{\mathbf{a}};\mathbf{f}) = \left\{ \begin{array}{ll}
P(\mathbf{a}'_M|\bar{\mathbf{a}};\mathbf{f}) & \textrm{if $\mathbf{a}'_{m} = \mathbf{a}_{m+1}$, for all $m = 1,\ldots,M-1$,}\\
0 & \textrm{otherwise.}
\end{array} \right.
\end{eqnarray*}

Let $\mathbf{v}_t(\mathbf{f})$ be the probability distribution on the state space
$\mathcal{A}^{M}$ in slot $t$ induced by protocol $\mathbf{f}$.
By the initialization in the definition of protocols with memory,
the initial distribution $\mathbf{v}_0(\mathbf{f})$
has element 1 for $\bar{\mathbf{a}} = (\mathbf{a}^0, \ldots, \mathbf{a}^0)$
and 0 elsewhere, where $\mathbf{a}^0$ denotes the idle outcome, $(W,\ldots,W)$.
Let $\mathbf{Q}(\mathbf{f})$ be the transition matrix of the Markov chain under
protocol $\mathbf{f}$. Then the probability distribution on $\mathcal{A}^{M}$ in slot $t$
can be computed by
\begin{align*}
\mathbf{v}_t(\mathbf{f}) = \mathbf{v}_0(\mathbf{f}) \mathbf{Q}(\mathbf{f})^t,
\end{align*}
for $t = 1,2,\ldots$.
Let $\mathcal{S}^i$ be the subset of $\mathcal{A}^{M}$ with user $i$'s success as the
most recent outcome, i.e., $\mathcal{S}^i \triangleq \{ \bar{\mathbf{a}} \in \mathcal{A}^{M}: \mathbf{a}_M = \mathbf{a}^i \}$.
Then the probability of user $i$'s success in slot $t$ is given by
$\tau_{i,t}(\mathbf{f}) \triangleq \sum_{\bar{\mathbf{a}} \in \mathcal{S}^{i}} v_t(\bar{\mathbf{a}}; \mathbf{f})$.
The fraction of slots with user $i$'s success,
or the throughput of user $i$, is given by
\begin{align*}
\tau_i(\mathbf{f}) \triangleq \lim_{J \rightarrow \infty} \frac{1}{J}
\sum_{t=1}^J \tau_{i,t}(\mathbf{f}),
\end{align*}
assuming that the limit exists.
If $\mathbf{f}$ is chosen so that the induced Markov chain has only
one closed communicating class, then there exists a unique stationary
distribution $\mathbf{v}(\mathbf{f})$, independent of
the initial distribution $\mathbf{v}_0(\mathbf{f})$, which satisfies
\begin{align} \label{eq:station}
\mathbf{v}(\mathbf{f}) = \mathbf{v}(\mathbf{f}) \mathbf{Q}(\mathbf{f}) \textrm{ and }
\mathbf{v}(\mathbf{f}) \mathbf{e} = 1,
\end{align}
where $\mathbf{e}$ is the column vector of length $|\mathcal{A}^{M}|$ whose elements are all 1 \cite{meyer}.
Then the expression for the throughput of user $i$ is reduced to
$\tau_{i}(\mathbf{f}) = \sum_{\bar{\mathbf{a}} \in \mathcal{S}^{i}} v(\bar{\mathbf{a}}; \mathbf{f})$.
Finally, the total throughput of the system under protocol
$\mathbf{f}$ is given by
\begin{align*}
\tau(\mathbf{f}) = \sum_{i=1}^N \tau_i(\mathbf{f}).
\end{align*}

\subsubsection{Average Delay}

The \emph{average delay} of a user is defined as the average
waiting time, measured in the unit of slots, until the beginning of its next
successful transmission starting from an arbitrarily chosen time.
Average delay under a protocol with memory $\mathbf{f}$ can be computed
using a Markov chain. Consider a slot to which an arbitrarily chosen time belongs,
and let $\bar{\mathbf{a}} \in \mathcal{A}^{M}$ be the outcomes in the
recent $M$ slots. Given
$\bar{\mathbf{a}}$, $\mathbf{f}$ yields a probability
distribution on the outcome in the next slot, $P(\mathbf{a}'|\bar{\mathbf{a}};\mathbf{f})$, given by \eqref{eq:trans23}.
Using $P(\mathbf{a}'|\bar{\mathbf{a}};\mathbf{f})$, we can compute the probability that user $i$ succeeds
for the first time after $r$ slots when
the outcomes in the recent $M$ slots are given by
$\bar{\mathbf{a}}$ and users follow $\mathbf{f}$,
\begin{align} \label{eq:nui}
\mu_i(r;\bar{\mathbf{a}},\mathbf{f}) \triangleq
Pr\{\mathbf{a}_{t + r} = \mathbf{a}^i \textrm{ and }
\mathbf{a}_{t + r'} \neq \mathbf{a}^i, \textrm{ for
all $r' = 1,\ldots,r-1$} | \bar{\mathbf{a}}_{t}=\bar{\mathbf{a}};\mathbf{f}\},
\end{align}
for $r = 1,2,\ldots$. For example,
\begin{align*}
\mu_i(1;\bar{\mathbf{a}},\mathbf{f}) &= P(\mathbf{a}_{t + 1}
= \mathbf{a}^i |
\bar{\mathbf{a}}_{t}=\bar{\mathbf{a}};\mathbf{f}) \textrm{ and}\\
\mu_i(2;\bar{\mathbf{a}},\mathbf{f}) &= \sum_{\mathbf{a} \in
\mathcal{A} \setminus \{\mathbf{a}^i\}} P(\mathbf{a}_{t + 2} =
\mathbf{a}^i | \bar{\mathbf{a}}_{t+1}=(\bar{\mathbf{a}}_{t}, \mathbf{a});\mathbf{f})
P(\mathbf{a}_{t + 1} =
\mathbf{a} | \bar{\mathbf{a}}_{t}=\bar{\mathbf{a}};\mathbf{f}),
\end{align*}
where $(\bar{\mathbf{a}}_{t}, \mathbf{a}) \in \mathcal{A}^{M}$ is
obtained by deleting the
first outcome in $\bar{\mathbf{a}}_{t}$ and adding $\mathbf{a}$
as the most recent outcome. Using \eqref{eq:nui}, we can compute the
average number of slots until user $i$'s next success starting from a slot with
the recent $M$ outcomes $\bar{\mathbf{a}}$:
\begin{align*}
d_i(\bar{\mathbf{a}};\mathbf{f}) \triangleq \sum_{r =
1}^{\infty} \mu_i(r;\bar{\mathbf{a}},\mathbf{f}) \, r.
\end{align*}
Since the distribution on the recent $M$ outcomes in slot $t$
is given by $\mathbf{v}_{t}(\mathbf{f})$, the average delay of user $i$
under protocol $\mathbf{f}$ can be computed as
\begin{align*}
D_i(\mathbf{f}) \triangleq \lim_{J \rightarrow \infty} \frac{1}{J}
\sum_{t=1}^J \sum_{\bar{\mathbf{a}} \in \mathcal{A}^{M}} v_t(\bar{\mathbf{a}}; \mathbf{f})
d_i(\bar{\mathbf{a}};\mathbf{f}) - 0.5,
\end{align*}
where $0.5$ is subtracted to take into account that the average time staying in the initial slot
is $0.5$ starting from an arbitrarily chosen time.
When the Markov chain has a unique stationary distribution $\mathbf{v}(\mathbf{f})$,
the expression for the average delay of user $i$ is reduced to
\begin{align*}
D_i(\mathbf{f}) = \sum_{\bar{\mathbf{a}} \in
\mathcal{A}^{M}} v(\bar{\mathbf{a}}; \mathbf{f})
d_i(\bar{\mathbf{a}};\mathbf{f}) - 0.5.
\end{align*}

\subsubsection{Discussion on Throughput and Average Delay}

The throughput of the system can be considered as an efficiency measure of
a protocol as it gives the channel utilization over time.
However, as its definition suggests, throughput reflects
the performance averaged over a long period of time and does not
contain much information about the short-term and medium-term performance.
As an illustration, consider two sequences of outcomes generated by repeating
$(\mathbf{a}^1, \mathbf{a}^2, \ldots, \mathbf{a}^N)$ and
$(\mathbf{a}^1,\mathbf{a}^1,\mathbf{a}^1,\mathbf{a}^1,\mathbf{a}^1, \mathbf{a}^2,
\mathbf{a}^2,\mathbf{a}^2,\mathbf{a}^2,\mathbf{a}^2, \ldots, \\ \mathbf{a}^N,
\mathbf{a}^N,\mathbf{a}^N,\mathbf{a}^N,\mathbf{a}^N)$.
In both sequences, each user succeeds in one out of $N$ slots
over a long period of time, as measured by throughput $1/N$. However, users may
prefer the first sequence to the second one as the first exhibits the steadier performance
over a short period of time. For example, suppose that a user counts its successes in $N$ consecutive slots
from an arbitrary slot. The first sequence guarantees one success in $N$ consecutive slots regardless
of the initial slot. On the contrary, in the second sequence, $N$ consecutive slots
contain no success with a high probability and many successes with a low probability. The variation in
the performance over a short period of time is captured by average delay.

A widely-used measure of delay in queueing theory is the inter-packet interval.
In our model, we can compute the \emph{average inter-packet time} of user $i$,
i.e., the average number of slots between two successes of user $i$, using
\begin{align} \label{eq:intpack}
\tilde{D}_i(\mathbf{f}) = \sum_{\bar{\mathbf{a}} \in
\mathcal{S}^{i}} \frac{v(\bar{\mathbf{a}}; \mathbf{f})}{\sum_{\bar{\mathbf{a}} \in
\mathcal{S}^{i}} v(\bar{\mathbf{a}}; \mathbf{f})} d_i(\bar{\mathbf{a}};\mathbf{f}).
\end{align}
We can interpret $d_i(\bar{\mathbf{a}};\mathbf{f})$ as the average time
to enter one of the states in $\mathcal{S}^{i}$ starting from the state
$\bar{\mathbf{a}}$ under protocol $\mathbf{f}$. Then $D_i(\mathbf{f})$
can be interpreted as the average time to enter $\mathcal{S}^{i}$ starting from
an arbitrary state, where the starting state is chosen following
the stationary distribution. Similarly, $\tilde{D}_i(\mathbf{f})$
can be interpreted as the average time to return $\mathcal{S}^{i}$ starting from
a state in $\mathcal{S}^{i}$.

Note that $d_i(\bar{\mathbf{a}};\mathbf{f})$ satisfies
\begin{align} \label{eq:avedel}
d_i(\bar{\mathbf{a}};\mathbf{f}) = 1 + \sum_{\bar{\mathbf{a}}' \in \overline{\mathcal{S}^{i}}}
Q(\bar{\mathbf{a}}'|\bar{\mathbf{a}},\mathbf{f}) d_i(\bar{\mathbf{a}}';\mathbf{f}),
\end{align}
for all $\bar{\mathbf{a}} \in \mathcal{A}^{M}$, where $\overline{\mathcal{S}^{i}}$ is
the complement of $\mathcal{S}^{i}$, i.e., $\overline{\mathcal{S}^{i}} \triangleq \mathcal{A}^{M}
\setminus \mathcal{S}^{i}$. Using \eqref{eq:station} and
\eqref{eq:avedel}, we can show that $\sum_{\bar{\mathbf{a}} \in
\mathcal{S}^{i}} v(\bar{\mathbf{a}}; \mathbf{f}) d_i(\bar{\mathbf{a}};\mathbf{f}) = 1$
for any protocol with memory $\mathbf{f}$, and thus \eqref{eq:intpack} can be rewritten as
\begin{align} \label{eq:inverse}
\tilde{D}_i(\mathbf{f}) = \frac{1}{\tau_i(\mathbf{f})},
\end{align}
which can be regarded as a version of Little's Theorem \cite{bert}.
Therefore, the average inter-packet time is completely
determined by throughput, and thus it provides no additional
information beyond that provided by throughput.

The waiting time paradox \cite{feller}
suggests that not only the mean of the inter-packet time but also
the variance of the inter-packet time matters for
the average waiting time measured from an arbitrarily chosen time.
Let $X_i$ be the random variable that represents the
inter-packet time of user $i$ with support $\mathbb{N} \triangleq \{1,2,\ldots\}$.
By definition, we have $E(X_i) = \tilde{D}_i$.
The Pollaczek-Khinchine (P-K) formula \cite{lazo} gives
the average residual time until the next success, or average delay,
by
\begin{align} \label{eq:PKformula}
D_i = \frac{E(X_i^2)}{2 E(X_i)} = \frac{1}{2} \left( 1+ \kappa_i^2 \right) \tilde{D}_i,
\end{align}
where $\kappa_{i}$ represents the coefficient of variation of $X_i$, i.e.,
$\kappa_i = \sqrt{Var(X_i)}/E(X_i)$. The P-K formula shows that average delay is increasing
in the variance of the inter-packet time for a given level of throughput
(or equivalently, for a given mean of the inter-packet time).

Consider the following three examples with a fixed level of the average
inter-packet time $\tilde{D}_i$ and different levels of the
coefficient of variation $\kappa_i$.
(i) Suppose that user $i$ has periodic successful transmissions.
Then it transmits a packet successfully once in $\tilde{D}_i$
slots, and its average delay is $\tilde{D}_i/2$ since $\kappa_{i} = 0$.
Note that $\tilde{D}_i/2$ is the smallest average delay
achievable with protocols that yield the average inter-packet time $\tilde{D}_i$.
(ii) Suppose that successes for user $i$ are
completely random in a sense that $X_i$ follows a geometric distribution.
Such a random variable can be generated by a memoryless protocol
that prescribes a fixed transmission probability for each user.
In this case, $\kappa_{i}^2 = 1 - 1/\tilde{D}_i$, and
thus $D_i = \tilde{D}_i - 0.5$.
(iii) Suppose that the successes of user $i$ are highly correlated over time.
Then $\kappa_{i} \gg 1$, and thus $D_i \gg \tilde{D}_i$. In this case,
user $i$ has frequent successes for a short period of time but sometimes
has to wait for a long period of time until its next success.

It is reasonable to assume that users prefer to have a steady
stream of transmissions as well as a high transmission rate.
As the above examples illustrate, the coefficient of variation of the
inter-packet time measures the volatility in successful transmissions
over time. Since average delay reflects the coefficient of variation of the
inter-packet time, we use it as a second performance metric to complement
the long-term performance metric, throughput.

\subsection{Problem of the Protocol Designer}

We formulate the problem faced by the protocol designer as
a two-stage procedure. First, the protocol designer chooses
the length of memory, $M$ slots, and the feedback technology, $\rho$.
Next, the protocol designer chooses a protocol from the
class of available protocols given the stage-one choice $(M,\rho)$.
For simplicity, we assume that the protocol designer considers only
symmetric protocols. For a symmetric protocol $f$, we have
$\tau_1(f) = \cdots = \tau_N(f)$ and $D_1(f) = \cdots = D_N(f)$.
Total throughput is given by $\tau(f) = N \tau_i(f)$ for any $i \in \mathcal{N}$, and
we use $D(f)$ to denote the average delay of each user.

The problem of the protocol
designer can be written formally as
\begin{align} \label{eq:oppd}
\max_{M \in \mathbb{N}, \rho \in \mathcal{R}} \left\{ \left( \max_{ f
\in \mathcal{F}_{M,\rho}} U(\tau(f), D(f)) \right) - C(M,\rho) \right\}.
\end{align}
$U$ represents the utility function of the protocol designer, defined
on total throughput and average delay. We assume that $U$ is
increasing in total throughput and decreasing in average delay
so that the protocol designer prefers a protocol that yields high
total throughput and small average delay. $C$ represents the cost
function of the protocol designer, defined on the length of memory
and the feedback technology. We assume that $C$ is increasing as
memory is longer and as the feedback technology is more informative.
Note that from a practical point of view the cost of expanding
memory is vanishingly small compared to the cost associated with
the feedback technology, which implies that the system is more
likely to be constrained by the available feedback technology
rather than by the size of memory. Hence, it is more natural to
interpret the cost associated with memory as the cost of implementing
protocols with a certain length of memory. For instance, the protocol designer
may prefer protocols with short memory to protocols with long memory because
the former is easier to program and validate than the latter.
We say that a protocol $f$ is \emph{optimal given $(M,\rho)$} if it
attains $\max_{ f \in \mathcal{F}_{M,\rho}} U(\tau(f), D(f))$.
We say that $f$ is an \emph{optimal protocol} for the protocol
designer if $f$ solves \eqref{eq:oppd}.

In order to analyze the performance of protocols with memory,
we approach the protocol designer's problem from two different
directions.
In Section IV, we first find the feasible throughput-delay pair
most preferred by the protocol designer. Then we show that the
most preferred throughput-delay pair can be achieved by a protocol
with memory.
In Section V, we focus on the simplest class of protocols with memory,
namely protocols with 1-slot memory, and investigate their properties
using numerical methods.

\section{Most Preferred Protocols}

Suppose that the protocol designer can choose
\emph{any} protocol at no cost. Then the protocol designer's problem
becomes
\begin{align} \label{eq:nocost}
\max_{ f \in \mathcal{F}} U(\tau(f), D(f)),
\end{align}
where $\mathcal{F}$ denotes the set of all symmetric protocols including those
that cannot be expressed as protocols with memory.
We say that $f$ is a \emph{most preferred protocol} if $f$ solves \eqref{eq:nocost}.
Consider a fixed level of total throughput $\tau$, which implies
individual throughput $\tau/N$ by the symmetry assumption on protocols.
Note that the P-K formula \eqref{eq:PKformula} was obtained without imposing
any structure on protocols, and it shows that for a given level of throughput,
average delay is minimized when there is no variation in the inter-packet
time of a user, i.e., when $\kappa_i = 0$. Hence, combining \eqref{eq:inverse}
and \eqref{eq:PKformula}, we can express the minimum average delay given total throughput $\tau$ as
\begin{align*}
D_{min}(\tau) = \frac{\tilde{D}_i}{2} = \frac{N}{2 \tau}.
\end{align*}
Since $D_{min}(\tau)$ is decreasing in $\tau$,
setting total throughput at the maximum level $1$
yields the minimum feasible average delay $D_{min} = D_{min}(1) = N/2$.
In other words, there is no protocol $f \in \mathcal{F}$ that attains
$D(f) < N/2$.
Hence, if there exists a protocol $f$ that achieves the maximum total
throughput 1 and the minimum average delay $N/2$ at the same time,
then $f$ is a most preferred protocol.

In order to obtain the most preferred throughput-delay pair $(1,N/2)$,
a protocol needs to provide each user with a successful transmission in every $N$
slots. TDMA is a protocol that achieves such a sequence of outcomes.
Since the labels of users are arbitrary,
we can describe the TDMA protocol as having user $i$ transmit in slot
$t$ if $(t\mod N) = (i\mod N)$ and wait in all other slots.
Then each user has one successful transmission in every $N$ slots, and thus
the TDMA protocol is a most preferred protocol. However, the TDMA
protocol requires coordination messages to be sent to users in order to assign time slots,
and thus it does not belong to the class of protocols with memory
defined in Section II.C. The following theorem establishes the existence
of a most preferred protocol in the class of protocols with memory.

\newtheorem{thm1}{Theorem}
\begin{thm1}
Assume $(N-1)$-slot memory and success/ failure (S/F) binary feedback $\rho_{SF}$.
Denote an $(N-1)$-slot history by $L = (a^1,z^1; \ldots; a^{N-1},z^{N-1})$, and
let $n(L)$ be the number of successes in $L$,
i.e., $n(L) = | \{ m: z^{m} = 1, m = 1,\ldots,N-1 \} |$.
Define a protocol with memory $f \in \mathcal{F}_{(N-1),\rho_{SF}}$ by
\begin{align*}
f(L) = \left\{
\begin{array}{ll}
0 \quad &\textrm{if $L$ contains $(T,1)$},\\
1/(N-n(L)) \quad &\textrm{otherwise}.
\end{array} \right.
\end{align*}
Then $\tau(f) = 1$ and $D(f) = N/2$.
\end{thm1}

\begin{IEEEproof}
Since $f(L) = 0$ if $L$ contains $(T,1)$, a user waits for $(N-1)$
slots following its success. Also, users that have no success in recent $(N-1)$
slots compete among themselves with the transmission probability
equal to the reciprocal of the number of such users. Suppose that
there is a success in each of the recent $(N-1)$ slots for the first time in
the system. Since a user waits for $(N-1)$ slots following its success, the $(N-1)$
successes must be by different users. Then the only user that has
not had a success in the recent $(N-1)$ slots transmits with probability 1
while other users wait in the current slot. From that point on,
a cycle of length $N$ containing one success of each user is repeated. Since
the probability of having $(N-1)$ consecutive successes before slot $t$ approaches 1
as $t$ goes to infinity, $f$ achieves a TDMA outcome
after a transient period with probability 1, and thus it is a most preferred protocol.
\end{IEEEproof}

Theorem 1 shows that with $(N-1)$-slot memory and S/F feedback,
$N$ users can determine their transmission slots in
a distributed way without need of explicit coordination messages,
thereby achieving the same outcome as TDMA after a transient period.
By trial and error, users find their transmission slots in a self-organizing
manner and stabilize to transmit once in every $N$ slots. Note that
$(N-1)$ is the minimum length of memory required to emulate
TDMA using a protocol with memory. A cycle of successes by
$N$ users cannot be generated with memory shorter than $(N-1)$ slots.
Also, S/F feedback is necessary to guarantee a success after
$(N-1)$ consecutive successes.

The expected duration of the transient period can be shortened
by using $N$-slot memory. Consider a protocol with $N$-slot memory,
$\tilde{f} \in \mathcal{F}_{N,\rho_{SF}}$, defined by
\begin{eqnarray} \label{eq:thm12}
\tilde{f}(L) = \left\{
\begin{array}{ll}
1 \quad &\textrm{if $(a^1,z^1) = (T,1)$},\\
0 \quad &\textrm{if $(a^1,z^1) = (W,1)$},\\
0 \quad &\textrm{if $z^1 \neq 1$ and $(a^m,z^m) = (T,1)$ for some $m = 2,\ldots,N$},\\
1/(N-n(L)) \quad &\textrm{if $z^1 \neq 1$ and $(a^m,z^m) \neq (T,1)$ for all $m = 2,\ldots,N$},
\end{array} \right.
\end{eqnarray}
where $n(L)$ is now the number of successes in recent $(N-1)$ slots.
The first and the third lines of \eqref{eq:thm12} state that once a user
succeeds, it waits for $(N-1)$ slots before the next transmission.
The second and the fourth lines state that a
user with no success in recent $N$ slots waits if the current
slot is already ``reserved'' by some user that succeeded $N$ slots ago
and contends with other such users if no user succeeded $N$ slots ago.
Hence, under
$\tilde{f}$, once a user succeeds in slot $t$, it is guaranteed to
succeed in slots $t+N$, $t+2N$, and so on, whereas under $f$ in Theorem 1 a user
that succeeded $N$ slots ago has to compete with other users with no
success in recent $(N-1)$ slots during the transient period.
Again, S/F feedback is necessary to let users know
whether the current slot is reserved or not. $\tilde{f}$ can be
considered as a generalization of reservation
Aloha with a frame of $N$ slots, where users can adjust their
transmission probabilities depending on the number of contending
users for non-reserved slots.

In practice, the number of users may vary over time as
users join and leave the system. When ternary feedback is available,
users can find the exact number of users in the system using the protocol $\tilde{f}$
as long as the number of users does not change too frequently. The initial estimate on the number of users
is set to be the maximum number of users that the system can allow.
When users follow $\tilde{f}$ based on their initial estimate,
there will be empty slots in a cycle obtained after a transient period.
Once a cycle is repeated, users can construct a new cycle by deleting the
empty slots, adjusting the number of slots in a frame equal to the actual number of users.
If some users leave the network, empty slots will appear in a frame,
and the remaining users can reduce the length of a frame
by deleting the empty slots. If some users join the network, we require them to transmit
immediately. Then a collision will occur, and it serves as a signal
to notify the existing users that there arrived a new user in the system.
Once a collision occurs after a transient period, users reset their
estimates to the maximum number of users and repeat the procedure
from the beginning in order to increase
the length of a frame by the number of new users.

\section{Protocols With 1-Slot Memory}

\subsection{Structure of Delay-Efficient Protocols}

We now examine protocols with 1-slot memory, which are the
simplest among protocols with memory.
Under protocols with 1-slot memory, users
determine their transmission probabilities using their
action-feedback pairs in the previous slot. In Appendix A, we explain
in detail how to compute throughput
and average delay under symmetric protocols with 1-slot memory using a Markov chain.
Since the protocol designer prefers a protocol with small average
delay for a given level of throughput, we consider the following
reduced problem of the protocol designer:
\begin{align} \label{eq:mindel}
D^*(\tau) \triangleq \min_{f \in \mathcal{F}_{1,\rho}} D(f)
\quad \textrm{subject to } \tau(f) = \tau,
\end{align}
for $\tau \in [0,1]$ and some feedback technology $\rho$. We say that a protocol $f$ is \emph{delay-efficient}
if it solves \eqref{eq:mindel} for $\tau = \tau(f)$, i.e., if $D(f) = D^*(\tau(f))$. Also, we
call the set of points $\{ (\tau,D^*(\tau)): \tau \in [0,1] \}$
the \emph{delay-efficiency boundary} of protocols with 1-slot memory. Since computing $\tau(f)$ and $D(f)$
for a given protocol $f \in \mathcal{F}_{1,\rho}$ involves solving matrix equations, it is in general difficult to
solve \eqref{eq:mindel} analytically, and thus we rely on numerical methods.

\begin{figure}%
\centering
\subfloat[][]{%
\label{fig:deleff-a}%
\includegraphics[width=0.45\textwidth]{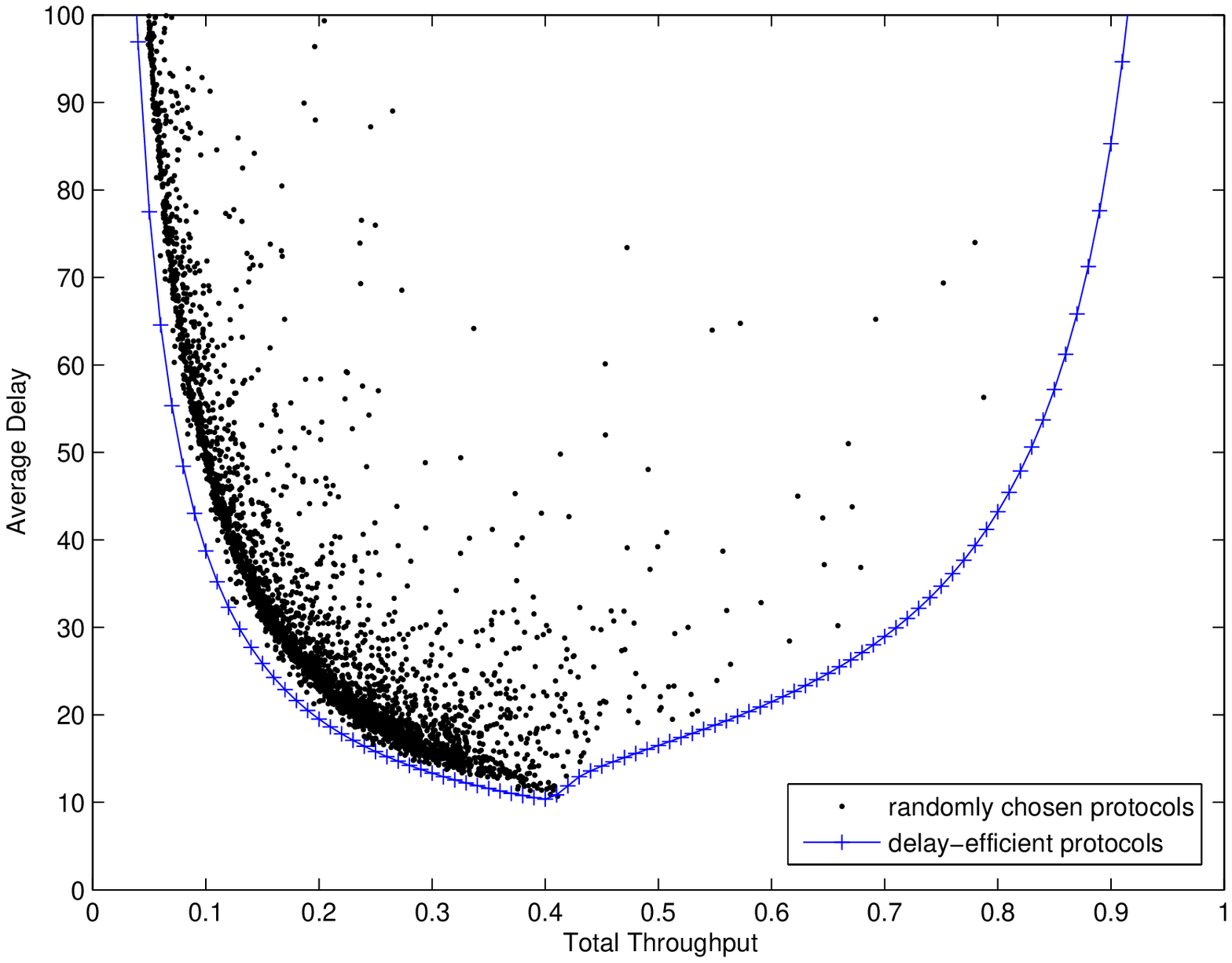}}%
\hspace{20pt}%
\subfloat[][]{%
\label{fig:deleff-b}%
\includegraphics[width=0.45\textwidth]{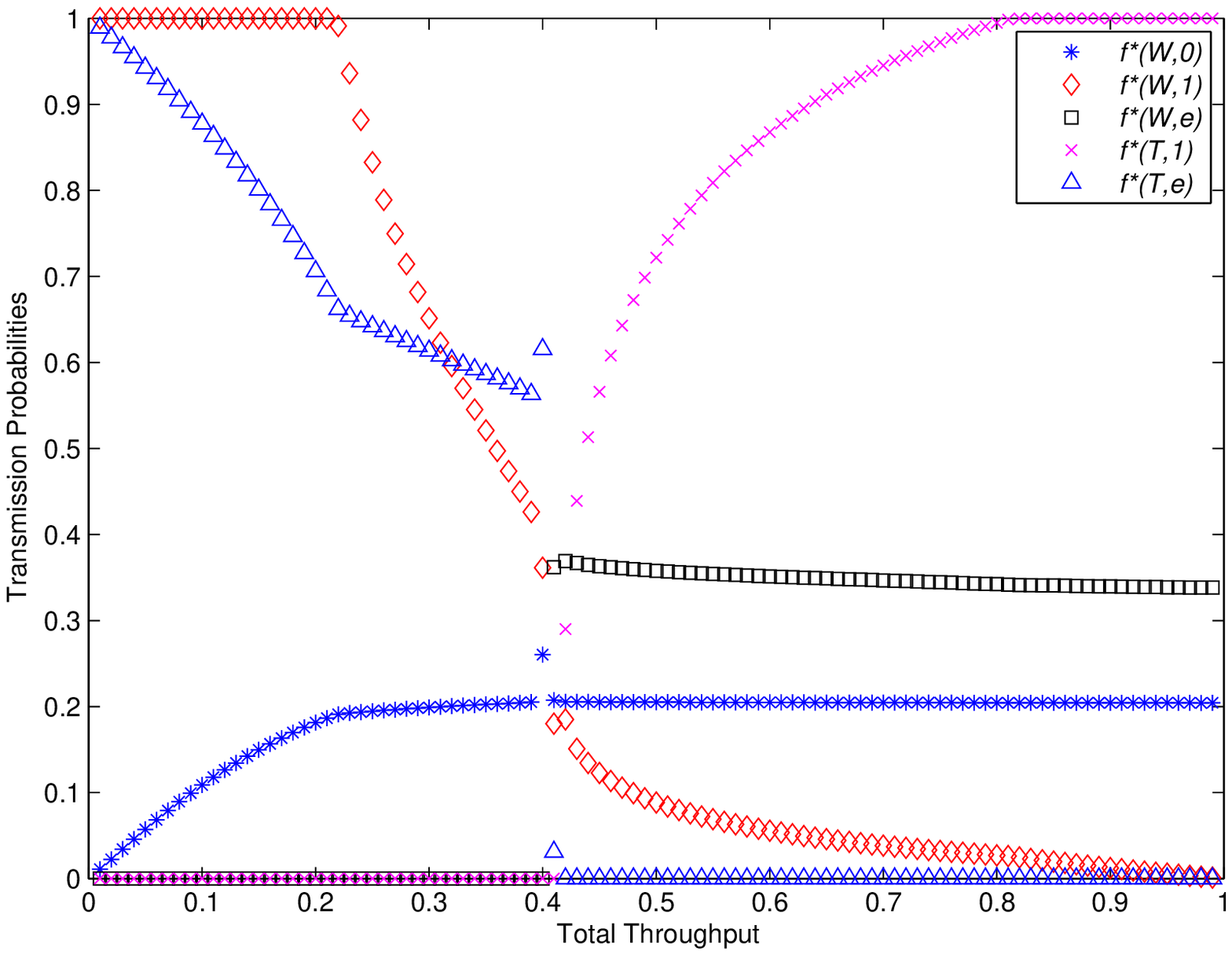}}%
\caption[Delay-efficient protocols.]{Delay-efficient protocols
with 1-slot memory:
\protect\subref{fig:deleff-a} total throughput and average delay under delay-efficient protocols and
randomly chosen protocols, and
\protect\subref{fig:deleff-b} transmission probabilities under delay-efficient protocols.}
\label{fig:deleff}%
\end{figure}

To obtain numerical results, we consider ternary feedback, denoted by $\rho_3$,
and five users, i.e., $N = 5$.
In order to guarantee the existence
of a stationary distribution for transition matrix $\mathbf{Q}(f)$, we restrict the range of $f$ (i.e., the set of
possible transmission probabilities) to be $[10^{-4}, 1-10^{-4}]$,
instead of $[0,1]$. We use the function \emph{fmincon} of MATLAB to
solve the restricted version of \eqref{eq:mindel}.\footnote{We vary
$\tau$ from 0.99 to 0.01 with a step size of $0.01$. We choose the initial protocol
for \emph{fmincon} as $(f(W,0), f(W,1), f(W,e), f(T,1), f(T,e)) = (1/N, 0, 1/(N-2), 1, 0)$ for $\tau = 0.99$
and use the solution for $\tau = n \times 0.01$ as the initial
protocol for $\tau = (n-1) \times 0.01$ from $n = 99$ to $n = 2$.}
Fig.~\ref{fig:deleff}\subref{fig:deleff-a} depicts
the delay-efficient boundary of protocols with 1-slot memory,
shown as a U-shaped curve.
Since the optimization problem to find
delay-efficient protocols is not necessarily convex, it is possible
that the numerical results locate local minima instead of global
minima. To validate the delay-efficiency of the protocols obtained
by the \emph{fmincon} function, we generate 5,000 symmetric protocols
with 1-slot memory where transmission probabilities are randomly chosen on $[0,1]$ and plot total throughput
and average delay under those protocols as dotted points in Fig.~\ref{fig:deleff}\subref{fig:deleff-a} (in the figure, only
3,631 points with average delay less than 100 are shown). The results in
Fig.~\ref{fig:deleff}\subref{fig:deleff-a} suggest that the numerically computed protocols are indeed (at least
approximately) delay-efficient.\footnote{Another interesting point to notice from Fig.~\ref{fig:deleff}\subref{fig:deleff-a}
is that most of the randomly chosen protocols yield
throughput-delay pairs close to those achieved by memoryless protocols. It suggests that
protocols with 1-slot memory need to be designed carefully in order to attain total throughput
above the maximum level achievable with memoryless protocols.}
Fig.~\ref{fig:deleff}\subref{fig:deleff-b} plots the transmission probabilities
under the delay-efficient protocols, denoted by $f^*$, as $\tau$ varies.
It shows that the structure of $f^*$ changes
around $\tau = 0.41$, which is also the turning point of the delay-efficiency boundary.
There is numerical instability for $\tau$ between 0.41 and 0.48 in that the solution depends
highly on the initial protocol. In fact, for $\tau$ in that region we can find protocols that
yield average delay slightly smaller than those in Fig.~\ref{fig:deleff}\subref{fig:deleff-a}
by specifying different initial protocols.

In the following, we provide an explanation for the shape of the delay-efficiency
boundary by investigating the structure of the delay-efficient
protocols.
In the low throughput region, ($\tau \leq 0.21$), the structure of
$f^*$ is given by
\begin{eqnarray*} 
f^*(W,0) \in [0, 0.2], \ f^*(W,1) = 1, \ f^*(W,e) = 0, \ f^*(T,1) = 0, \ f^*(T,e) \in [0.65, 1].
\end{eqnarray*}
Since $f^*(T,1) = 0$ and $f^*(W,1) = 1$, a success lasts for only
one slot and is followed by a collision. Since $f^*(T,e)$ is high,
a collision is likely to be followed by another collision.
Similarly, an idle slot is likely to be followed by another idle
slot since $f^*(W,0)$ is low. This means that total throughput is
kept low by inducing many idle or collision slots between two
successes. Hence, as total throughput increases in the low
throughput region, the expected number of idle or collision slots between two
successes is reduced, and as a result users transmit their packets
more frequently. According to the P-K formula \eqref{eq:PKformula}, average delay
is determined by the mean and the coefficient of variation of the inter-packet time.
On the left-hand side of the delay-efficiency boundary ($\tau \leq 0.4$),
the average inter-packet time is reduced
while the coefficient of variation of the inter-packet time remains about the same
as $\tau$ increases, resulting in the inverse relationship between
throughput and average delay.

In the high total throughput region ($\tau \geq 0.82$), the
structure of $f^*$ is given by
\begin{eqnarray} \label{eq:deprot}
f^*(W,0) \approx \frac{1}{N}, \ f^*(W,1) \approx 0, \ f^*(W,e)
\approx \frac{1}{N-2}, \ f^*(T,1) = 1, \ f^*(T,e) = 0.
\end{eqnarray}
In a slot following an idle slot, users transmit with probability close to $1/N$.
In a slot following a success, the successful user
transmits with probability 1 while other users wait with high probability. The
transmission probability of other users approaches 0 as $\tau$
becomes close to 1. In a slot following a collision, users that transmitted in
the collision wait while other users transmit with probability close to
$1/(N-2)$. Since a collision involving two transmissions is most
likely among all kinds of collisions, setting $f^*(W,e)=1/(N-2)$ maximizes
the probability of success given that colliding users wait.
$f^*(T,1) = 1$ and $f^*(W,1) \approx 0$ is the key feature of
protocols with 1-slot memory that achieve high total throughput. By
correlating successful users in two consecutive slots, protocols with
1-slot memory can yield total throughput arbitrarily close to 1. However, higher
throughput is achieved by allowing a successful user to
use the channel for a longer period, which makes other users wait
longer until they transmit next time.
On the right-hand side of the delay-efficiency boundary ($\tau > 0.4$),
the coefficient of variation of the inter-packet time increases without bound
as $\tau$ approaches 1, resulting in the positive relationship between
throughput and average delay.

\begin{figure}
\centering
\includegraphics[width=0.6\textwidth]{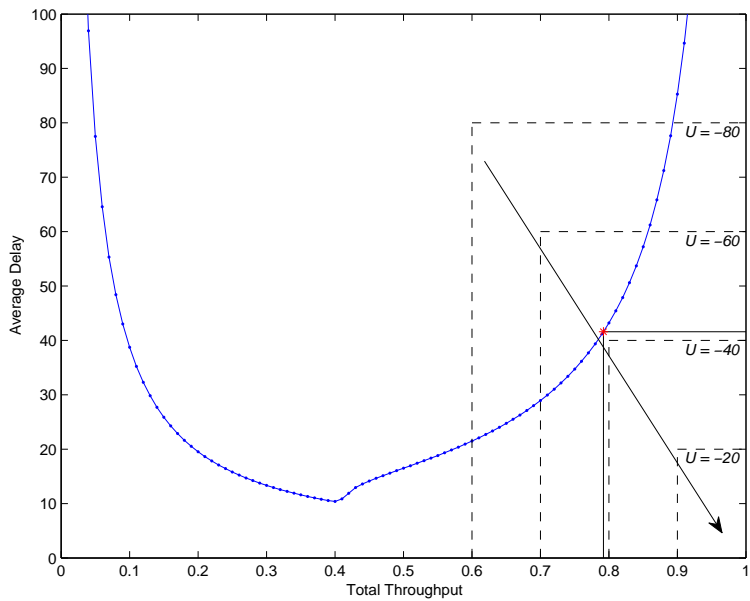}
\caption{Illustration of an optimal protocol.}
\label{fig:utilopt}
\end{figure}

Since the protocol designer prefers a protocol that yields high
total throughput for a given level of average delay, optimal protocols
must lie on the right-hand side of the delay-efficiency boundary.
As an illustrative example, suppose that the utility
function of the protocol designer is given by
\begin{align*} 
U(\tau,D) = - \max \{200(1-\tau),D\}.
\end{align*}
Then an increase in total throughput by 0.1 has the same utility
consequence as a decrease in average delay by 20 slots. In Fig.~\ref{fig:utilopt},
dashed curves depict the indifference curves of the utility
function, each of which represents the throughput-delay pairs that
yield the same level of utility, while the
arrow shows the increasing direction of the utility function. The
throughput-delay pair that
maximizes the utility of the protocol designer is $(0.792, 41.6)$,
marked with an asterisk in Fig.~\ref{fig:utilopt}. The protocol designer can
determine the optimal protocol $f^{o}$ given $(M,\rho) =
(1,\rho_3)$ by finding a protocol $f \in \mathcal{F}_{1,\rho_3}$
that yields $(\tau(f), D(f)) = (0.792, 41.6)$,
which is
\begin{eqnarray} \label{eq:uoprot}
f^{o}(W,0) = 0.20, \ f^{o}(W,1) = 0.03, \ f^{o}(W,e) = 0.34, \ f^{o}(T,1) = 0.99, \ f^{o}(T,e) = 0.
\end{eqnarray}

\subsection{Robustness Properties of Delay-Efficient Protocols}

\subsubsection{Unknown Number of Users}

\begin{figure}
\centering
\includegraphics[width=0.6\textwidth]{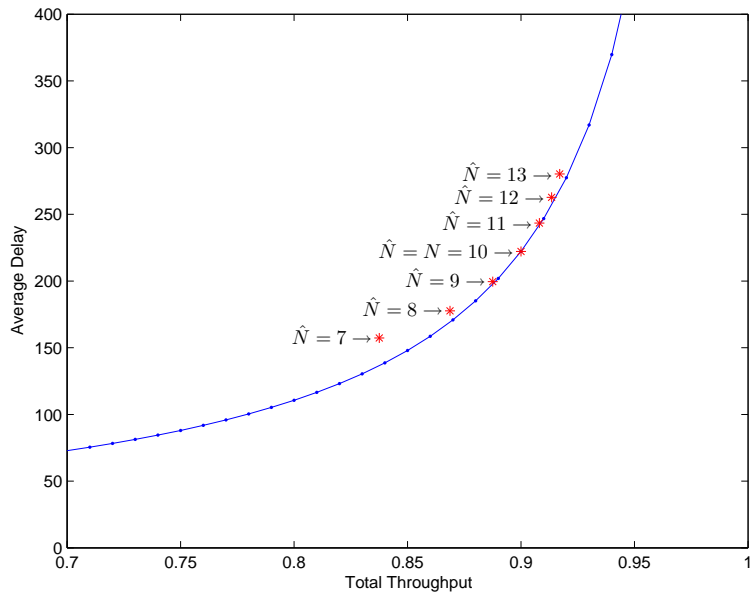}
\caption{Total throughput and average delay under
delay-efficient protocols based on the estimated numbers of users when
the actual number of users is ten.}
\label{fig:robust}
\end{figure}

We relax the assumption that users know the exact number of users in
the system. Instead, we assume that each user has an estimate on the
number of users and executes the prescribed protocol based on the estimate.%
\footnote{In other words, the protocol designer specifies a protocol as a function of the numbers of users,
not a fixed protocol.} We consider a scenario where there are ten
users and ternary feedback is available.
For simplicity, we assume that the objective of the protocol
designer is to achieve total throughput 0.9 and
that all users have
the same estimate on the number of users, which will be the case
when users use the same estimation method based on past channel feedback.
Fig.~\ref{fig:robust} shows a portion of the delay-efficiency boundary
with $N = 10$ and plots the throughput-delay pairs when the ten users
follow delay-efficient protocols that are computed to achieve total throughput 0.9
based on the estimated number of users, denoted by $\hat{N}$, between 7 and 13.
The results from Fig.~\ref{fig:robust}
suggest the robustness of delay-efficient protocols with respect to variations in the
number of users in a sense that as the estimated number of users is close
to the actual one, the protocol designer obtains a performance
close to the desired one.

Based on the observations from Fig.~\ref{fig:robust}, we can
consider the following procedure to dynamically adjust the estimates
of users. Users update their estimates periodically by comparing
the actual total throughput since the last update with the desired
total throughput.\footnote{Note that users can compute the actual total throughput using ternary feedback.}
Users increase (resp. decrease) their estimates by one
if the actual total throughput is lower
(resp. higher) than the desired total throughput
by a certain threshold level.
When designed carefully, this estimation procedure will make the estimated number of users converge to
the actual number of users because the actual total throughput
is lower than (resp. equal to, higher than) the desired total throughput if the estimated number of users
is smaller than (resp. equal to, larger than) the actual number.
This procedure can be regarded as an extension of the pseudo-Bayesian algorithm of
\cite{riv} in which users adjust their estimates in every slot based on
the channel feedback of the previous slot. In the proposed estimation
procedure, memory is utilized not only to coordinate transmissions
but also to estimate the number of users.

\subsubsection{Errors in Feedback Information}

So far we have restricted our attention to deterministic feedback
technologies. We relax this restriction and consider stochastic
feedback technologies. As discussed in Section II.B., stochastic
channel feedback for user $i$ can be represented by a mapping $h_i$
from $\mathcal{A}$ to $\triangle(H_i)$. If the structure of channel
feedback, $(h_1, \ldots, h_N)$, is known to the protocol designer,
it can be modeled in the transition matrix $\mathbf{Q}(f)$ in Section III
by extending the state space from $\mathcal{A}^M$ to $(\mathcal{L}_{\rho}^N)^M$.
Then the protocol designer can find an optimal protocol
taking into account randomness in feedback information.

Here we introduce random errors in channel feedback, which are not
modeled by the protocol designer, and examine the
performance of an optimal protocol in the presence of random
errors. We assume that ternary feedback is available but subject to
random errors. In particular, a user obtains the correct feedback signal
with probability $1 - 2 \epsilon$ and each of the two incorrect signals
with probability $\epsilon$, for a small $\epsilon > 0$. For example, if there is no
transmission in the system, then a user receives feedback $0$ with probability $1 -
2 \epsilon$ and each of feedback $1$ and $e$ with probability $\epsilon$.
We assume that the feedback signals of
users are independent. We continue to assume that ACK feedback is
perfect, and thus transmitting users always learn the correct results of
their transmission attempts, regardless of the realization of channel feedback.

\begin{table}
\caption{Total throughput and average delay with a stochastic
feedback technology.}
\centering
\begin{tabular}{|c|c|c|c|}
\hline
\multicolumn{2}{|c|}{} & Total throughput & Average delay \\
\hline \hline \multicolumn{2}{|c|}{Analysis} & 0.7920 & 41.5935 \\
\hline \multirow{8}{*}{Simulation} & $\epsilon = 0$ & 0.7910 & 41.2375 \\ \cline{2-4}
& $\epsilon = 0.01$ & 0.7667 & 37.4377 \\ \cline{2-4}
& $\epsilon = 0.02$ & 0.7441 & 33.4907 \\ \cline{2-4}
& $\epsilon = 0.03$ & 0.7235 & 31.4114 \\ \cline{2-4}
& $\epsilon = 0.05$ & 0.6844 & 28.0600 \\ \cline{2-4}
& $\epsilon = 0.07$ & 0.6467 & 25.2149 \\ \cline{2-4}
& $\epsilon = 0.10$ & 0.6049 & 22.9282 \\ \cline{2-4}
& $\epsilon = 0.20$ & 0.4996 & 19.0503 \\
\hline
\end{tabular}
\label{table:errors}
\end{table}

Table \ref{table:errors} shows the performance of the
optimal protocol $f^{o}$ in \eqref{eq:uoprot} at the various levels of
$\epsilon$ when $N = 5$. To obtain the simulation results,
we generate transmission decisions and feedback information for $100,000$ slots,
for each level of $\epsilon$.
Table \ref{table:errors} suggests that delay-efficient protocols
have a robustness (or continuity) property with respect to
random errors in feedback information, since the obtained performance is close to the
desired one when the error level is small.
Note that an error occurring to a waiting user following a success
induces the user to transmit with a higher probability (i.e., to transmit
with probability $f^o(W,0)$ or $f^o(W,e)$ instead of $f^o(W,1)$), making consecutive successes
last shorter. Thus, as the error level increases, both total throughput and
average delay decrease. The obtained throughput-delay pairs remain
close to the delay-efficiency boundary, suggesting that errors following
an idle or a collision slot cause little performance degradation.

\subsection{Comparison of Channel Feedback}

\begin{figure}
\begin{center}
\includegraphics[width=0.6\textwidth]{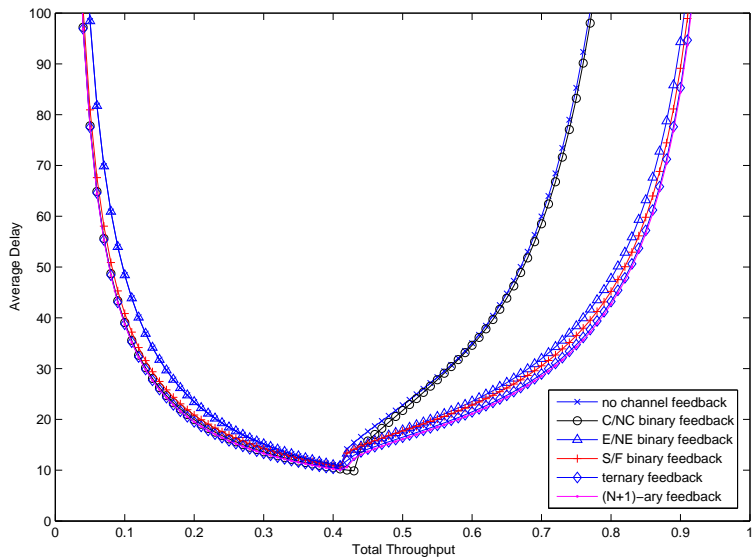}
\caption{Delay-efficiency boundaries under various channel feedback.}
\label{fig:valueFT}
\end{center}
\end{figure}

We now analyze the impact of the different forms of channel feedback on the performance
of protocols with 1-slot memory. As mentioned in Section II.C., the set of available
protocols $\mathcal{F}_{M,\rho}$ expands as the feedback technology $\rho$
becomes more informative (in other words, a protocol $f \in \mathcal{F}_{M,\rho}$ can always
be replicated by another protocol $f' \in \mathcal{F}_{M,\rho'}$ if $\rho'$ is
more informative than $\rho$). This relationship implies that a more
informative feedback technology yields a lower delay-efficiency boundary for a given length of memory.
We consider six feedback technologies with ACK feedback and different channel feedback models:
no channel feedback, S/F binary feedback, C/NC binary feedback, E/NE binary feedback,
ternary feedback, and $(N+1)$-ary feedback, as introduced in Section II.B.
Fig.~\ref{fig:valueFT} depicts the delay-efficiency boundaries of
protocols with 1-slot memory under the six feedback technologies. As
expected, for a given level of throughput, the minimum average delay becomes
smaller as we move from no feedback to binary feedback, to ternary feedback,
and to $(N+1)$-ary feedback.\footnote{In Fig.~\ref{fig:valueFT}, average delay
is smallest under C/NC binary feedback for $\tau = 0.42$ and $0.43$, which
is a result of numerical instability around the turning point of the delay-efficiency
boundaries, as pointed out in Section V.A.}

Fig.~\ref{fig:valueFT} shows that in the operating region of the protocol
designer (i.e., the right-hand side of the delay-efficiency boundaries),
protocols with 1-slot memory perform worse under no channel feedback and
C/NC binary feedback than under other considered channel feedback.
In order to obtain high throughput with 1-slot memory, we
need a high correlation between successful users in two consecutive slots, which
requires $f(W,1) \approx 0$ and $f(T,1) \approx 1$. However, under no channel
feedback or C/NC binary feedback, a waiting user cannot distinguish
between idle slots and success slots, leading to $f(W,0) = f(W,1) \approx 0$.
This makes idle slots last long once one occurs, resulting in large average
delay.
Under S/F binary feedback, a user is constrained to use $f(W,0) = f(W,e)$.
As can be seen in Fig.~\ref{fig:deleff}\subref{fig:deleff-b}, the values of
$f(W,0)$ and $f(W,e)$ are not much different in delay-efficient protocols
under ternary feedback. Thus, using a single probability $f(W,0 \cup e)$ instead
of two different probabilities, $f(W,0)$ and $f(W,e)$, causes only minor performance
degradation.
Under E/NE binary feedback, a delay-efficient protocol with 1-slot memory
that achieves high throughput has the following structure:
\begin{eqnarray} \label{eq:sfbin}
f(W,0) \approx \frac{1}{N}, \ f(W,1 \cup e) = 0, \ f(T,1)
\approx 1, \ f(T,e) \approx \frac{1}{2}.
\end{eqnarray}
Following a collision, only colliding users transmit under \eqref{eq:sfbin}
whereas only noncolliding users transmit under \eqref{eq:deprot}.
Fig.~\ref{fig:valueFT} shows that the restriction that E/NE binary feedback imposes compared
to ternary feedback has a small impact on performance.
Fig.~\ref{fig:valueFT} also shows that the improvement in performance
from having $(N+1)$-ary feedback over ternary feedback is only marginal.
The ability of users to distinguish
the exact numbers of transmissions in collisions does not help much
because collisions involving three or more transmissions rarely occur under
ternary feedback in the high throughput region.

\subsection{Comparison of Protocols}

\begin{figure}
\begin{center}
\includegraphics[width=0.6\textwidth]{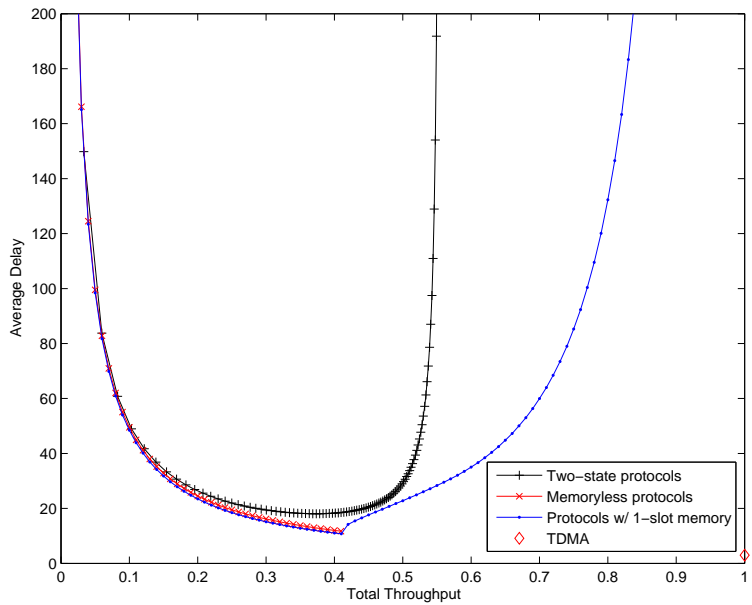}
\caption{Total throughput and average delay under two-state
protocols, memoryless protocols, protocols with 1-slot memory, and TDMA.}
\label{fig:comparison}
\end{center}
\end{figure}

We consider $N = 5$ and compare the
performance of four different kinds of protocols: protocols with 1-slot memory,
two-state protocols in \cite{ma}, memoryless protocols, and TDMA.
In Fig.~\ref{fig:comparison}, the delay-efficiency boundary of protocols
with 1-slot memory is shown assuming no channel feedback (i.e., ACK feedback only)
because two-state protocols can be implemented without using channel feedback.
\cite{ma} proposes two-state protocols of the form
$p_F = 1$ and $p_G = 1 - \sqrt[N-1]{1 - 1/\eta}$, where
$\eta$ is called a short-term fairness parameter, which measures the
average duration of consecutive successes. We vary $1/\eta$ from
0.01 to 1 with a step size of 0.01 to generate the throughput-delay pairs under
two-state protocols plotted in Fig.~\ref{fig:comparison}.
The result confirms that total throughput
achievable with a two-state protocol is bounded from above by
$N/(2N-1) = 5/9$, whereas protocols with 1-slot memory can attain
any level of total throughput between 0 and 1. Moreover, in the range of
total throughput achievable with a two-state
protocol, delay-efficient protocols with 1-slot memory yield smaller average
delay than two-state protocols. Since
two-state protocols with $p_F = 1$ can be considered as imposing
a restriction of $f(W,\emptyset) = f(T,e)$, these results suggest that there is
a significant performance degradation by assigning the same transmission
probability following the two action-feedback pairs, $(W,\emptyset)$ and $(T,e)$.
Moreover, by combining Fig.~\ref{fig:comparison} with Fig.~\ref{fig:valueFT},
we can see that the performance degradation from using a two-state
protocol instead of a protocol with 1-slot memory is severer when channel
feedback is available.

A symmetric memoryless protocol can be represented by a single transmission
probability, which is used regardless of past histories. The throughput
and the average delay of user $i$ under memoryless protocol $f \in [0,1]$
are given by $\tau_i(f) = f(1-f)^{N-1}$ and $D(f) = 1/f(1-f)^{N-1} - 0.5$, respectively.
The optimal memoryless protocol is thus the transmission probability
that maximizes $\tau_i(f)$ and minimizes $D(f)$ at the same time,
which is given by $f_0 = 1/N$ \cite{massey}. The throughput-delay pairs under
memoryless protocols in Fig.~\ref{fig:comparison} are obtained by
varying $f$ from 0 to 1. The lower-right point of the throughput-delay curve
corresponds to the throughput-delay pair under the optimal
memoryless protocol, $(\tau(f_0), D(f_0)) = ((1 - 1/N)^{N-1},
N/(1 - 1/N)^{N-1}-0.5) = (0.41, 11.71)$.
Protocols with 1-slot memory also outperform memoryless protocols in that protocols with 1-slot memory
support a wider range of achievable total throughput than memoryless protocols do.
In Section IV, we have seen that TDMA
achieves the most preferred throughput-delay pair, $(1,N/2) = (1,2.5)$.
Protocols with 1-slot memory cannot achieve
the most preferred throughput-delay pair because memory of length at least $(N-1)$ slots
is necessary to obtain it.

\section{Application to Wireless Local Area Networks}

In the idealized slotted multiaccess
system considered so far, all packets are of
equal size, the transmission of a packet takes the duration of one
slot, and users receive immediate feedback information. We relax these assumptions to
apply protocols with memory to WLANs. In particular, packet sizes may
differ across packets, and we take into consideration propagation and detection
delay as well as overhead such as a packet header and an ACK signal.
We consider a WLAN model where users follow a random
access scheme using transmission probabilities based on CSMA/CA.
In the WLAN model, the duration of slots
depends on the channel state (idle, success, or collision). Let
$\sigma_0$, $\sigma_1$, and $\sigma_2$ be the duration of a slot
when the channel state is idle, success, and collision,
respectively.
The expressions for $\sigma_1$ and $\sigma_2$ can be
found in (14) and (17) of \cite{bianchi}, depending on
whether the RTS/CTS mechanism is disabled or not. Total throughput
is expressed as
\begin{align} \label{eq:dcf}
\tau = \frac{P_1 E[P]}{P_0 \sigma_0 + P_1 \sigma_1 + P_2 \sigma_2},
\end{align}
where $E[P]$ is the average packet transmission duration and $P_0$,
$P_1$, and $P_2$ are the fractions of idle, success, and
collision slots, respectively. In the idealized slotted model, we
assume that the size of each packet is equal to the slot duration
and ignore overhead so that $\sigma_0 = \sigma_1 = \sigma_2 = E[P]$,
and thus the expression for total throughput in \eqref{eq:dcf} is reduced to $P_1$,
the fraction of success slots. \cite{bianchi} shows that the IEEE
802.11 DCF protocol can be approximated by a protocol that prescribes
a single transmission probability, i.e., a memoryless protocol.
The transmission probability corresponding to DCF is determined as a
function of the minimum and maximum contention window sizes by
solving (7) and (9) of \cite{bianchi} simultaneously. In
Appendix B, we derive the expressions for throughput and
average delay under symmetric protocols with 1-slot memory and
memoryless protocols in the WLAN model.

\begin{figure}
\begin{center}
\includegraphics[width=0.6\textwidth]{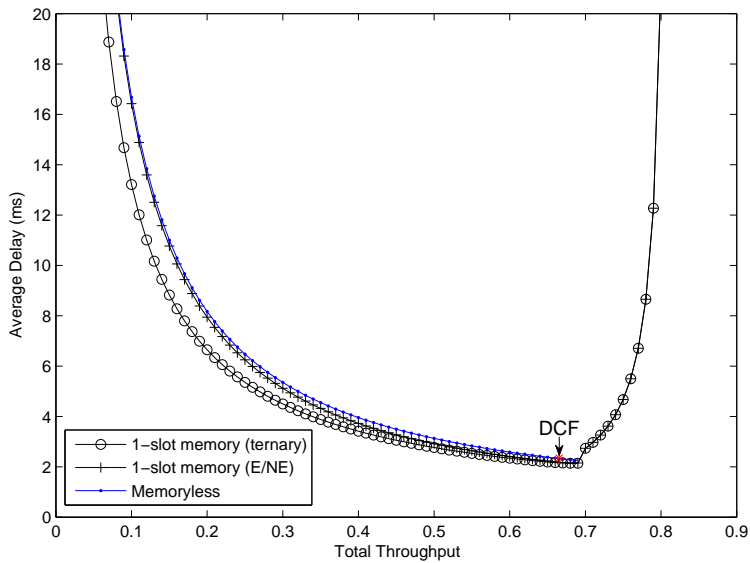}
\caption{Total throughput and average delay under delay-efficient
protocols with 1-slot memory and memoryless protocols in the WLAN model.}
\label{fig:DCF}
\end{center}
\end{figure}

Fig.~\ref{fig:DCF} depicts the delay-efficiency boundaries of protocols
with 1-slot memory under ternary feedback and E/NE binary feedback. It also
plots the throughput-delay pairs under memoryless protocols as well as
the throughput-delay pair achieved by the memoryless protocol corresponding to DCF.
To obtain numerical results, we
consider $N = 5$ and use parameters
specified by IEEE 802.11a PHY mode-8 \cite{ieee}, which are
tabulated in Table~\ref{table:DCF}. The values of $E[P]$, $\sigma_0$,
$\sigma_1$, and $\sigma_2$ are 341.33, 9, 419.56 and 400.48,
respectively, in $\mu$s. To find the transmission probability
corresponding to DCF, we use the minimum and maximum contention
window sizes $CW_{min} = 16$ and $CW_{max} = 1024$.
An upper bound on total throughput can be obtained by setting $P_1 = 1$,
which yields $\bar{\tau} = E[P]/\sigma_1 \approx 0.8136$. We compute
delay-efficient protocols with 1-slot memory for $\tau$ between 0.01 and 0.81 with
a step size of 0.01 using the same numerical method as in Section V.

Comparing Fig.~\ref{fig:DCF} with Fig.~\ref{fig:comparison},
we can see that by utilizing a carrier sensing scheme,
memoryless protocols resolve contention more efficiently in the
WLAN model than in the slotted multiaccess model.
Since idle slots are much shorter than collision
slots in WLAN, idle slots have much smaller effects
on total throughput and average delay than collision slots
have. Thus, in WLAN, transmission probabilities can be set low
to achieve a success without experiencing many collisions.
Carrier sensing also induces the turning point of the delay-efficiency boundaries in WLAN
to occur around $\tau = 0.69$, yielding the narrower and steeper right-hand side of
the delay-efficiency boundaries. The main findings in Section V.D. that protocols
with 1-slot memory can achieve smaller average delay
for a given level of total throughput and a wider range of total throughput
compared to memoryless protocols
remain valid in the WLAN model.
On the right-hand side of the delay-efficiency boundaries, i.e., for $\tau \geq 0.7$,
the delay-efficient protocols with 1-slot memory under ternary feedback have
the structure of $f(W,1) = f(W,e) = 0$, and thus
the performance is not affected by having E/NE binary feedback
instead of ternary feedback.

\begin{table}
\caption{IEEE 802.11a PHY mode-8 parameters.} \centering
\begin{tabular}{|c|c|}
\hline
Parameters & Values \\
\hline \hline
Packet payload & 2304 octets \\
MAC header & 28 octets \\
ACK frame size & 14 octets \\
\hline
Data rate & 54 Mbps \\
Propagation delay & 1 $\mu$s \\
Slot time & 9 $\mu$s \\
PHY header time & 20 $\mu$s \\
SIFS & 16 $\mu$s \\
DIFS & 34 $\mu$s \\
\hline
\end{tabular}
\label{table:DCF}
\end{table}

\section{Conclusion}

In this paper, we have investigated how memory can be utilized
in MAC protocols to achieve coordination without relying on explicit
control messages.
With $(N -1)$-slot memory, $N$ users can share the channel
as in TDMA.
With 1-slot memory, high throughput can be obtained
by correlating successful users in two consecutive slots,
which results in large average delay.
Generalizing these results, with $M$-slot memory, where $M \leq N-2$,
we can have the first $M$ successful users use the channel alternatingly
while a collision created by a non-successful user with a small probability
leads to a potential change of hands for the collision slot.

Our framework can be extended in several directions. First,
we can consider asymmetric protocols to provide quality of service
differentiation across users. Second, as the literature on
repeated games suggests, memory can also be used to sustain
cooperation among selfish users. By utilizing memory,
users can monitor the behavior of other users and punish
misbehavior. Lastly, the basic idea of this paper can be carried over to
a general multi-agent scenario where it is desirable to have one agent
behave in a different way than others.

\appendices

\section{Derivation of Throughput and Average Delay
Under Symmetric Protocols With 1-Slot Memory in the Slotted Multiaccess System}

Exploiting the symmetry of protocols, we can consider a Markov chain
for a representative user whose state is defined as a pair of its
transmission action and the number of transmissions, instead of
a Markov chain with the state space $\mathcal{A}$ as in Section III.
Let user $i$ be the representative user. The state of user $i$ in the current
slot when the outcome of the previous slot was $\mathbf{a}$ is given by $s = (a_i, k(\mathbf{a}))$.
We use $\mathcal{S}$ to denote the set of all states, as defined in Section II.B.
There are total $2N$ states, which we list as
\begin{align} \label{eq:list}
(T,1), \ldots, (T,N), (W,0), \ldots, (W,N-1).
\end{align}
Suppose that users follow a symmetric protocol $f \in \mathcal{F}_{1,\rho}$. Then we can express the
transition probabilities across states in terms of $f$. If
user $i$ is in state $(T,k)$, then
$k$ users including user $i$
transmit with probability $f(T,\rho(T,k))$, and $(N-k)$ users with
probability $f(W,\rho(W,k))$. Hence, a
transition from $(T,k)$ to $(a',k')$ occurs with probability
\begin{align} \label{eq:q1}
&Q((T,k')|(T,k);f) = f(T,\rho(T,k))
\sum_{\{(x,y) \in \mathbb{N}_0^2 : x+y = k'-1, \, x \leq
k-1, \, y \leq N-k \}} \\ &\left[ \binom{k-1}{x} f(T,\rho(T,k))^x (1 -
f(T,\rho(T,k)))^{k-1-x} \binom{N-k}{y}
f(W,\rho(W,k))^y (1 - f(W,\rho(W,k)))^{N-k-y} \right] \nonumber
\end{align}
for $k' = 1,\ldots,N$, and
\begin{align} \label{eq:q2}
&Q((W,k')|(T,k);f) = (1-f(T,\rho(T,k))) \sum_{\{(x,y) \in \mathbb{N}_0^2 : x+y = k', \, x \leq k-1, \, y
\leq N-k \}} \\ &\left[ \binom{k-1}{x} f(T,\rho(T,k))^x (1
- f(T,\rho(T,k)))^{k-1-x} \binom{N-k}{y}
f(W,\rho(W,k))^y (1 - f(W,\rho(W,k)))^{N-k-y} \right] \nonumber
\end{align}
for $k' = 0,\ldots,N-1$, where $\mathbb{N}_0$ is the set of
nonnegative integers. Similarly, if user $i$ is in state $(W,k)$,
then $k$ users transmit with probability
$f(T,\rho(T,k))$, and $(N-k)$ users including user $i$ with probability
$f(W,\rho(W,k))$. A transition from $(W,k)$ to
$(a',k')$ occurs with probability
\begin{align} \label{eq:q3}
&Q((T,k')|(W,k);f) = f(W,\rho(W,k)) \sum_{\{(x,y) \in \mathbb{N}_+^2 : x+y = k'-1, \, x \leq k, \, y
\leq N-k-1 \}} \\ &\left[ \binom{k}{x}
f(T,\rho(T,k))^x (1 - f(T,\rho(T,k)))^{k-x} \binom{N-k-1}{y} f(W,\rho(W,k))^y (1 - f(W,\rho(W,k)))^{N-k-1-y}
\right] \nonumber
\end{align}
for $k' = 1,\ldots,N$, and
\begin{align} \label{eq:q4}
&Q((W,k')|(W,k);f) = (1-f(W,\rho(W,k))) \sum_{\{(x,y) \in \mathbb{N}_+^2 : x+y = k', \, x \leq k, \, y \leq
N-k-1 \}} \\ &\left[ \binom{k}{x}
f(T,\rho(T,k))^x (1 - f(T,\rho(T,k)))^{k-x} \binom{N-k-1}{y} f(W,\rho(W,k))^y (1 - f(W,\rho(W,k)))^{N-k-1-y}
\right] \nonumber
\end{align}
for $k' = 0,\ldots,N-1$.

Using \eqref{eq:q1}--\eqref{eq:q4}, we obtain a $(2N \times 2N)$
matrix $\mathbf{Q}(f)$. The $(j,j')$-entry of $\mathbf{Q}(f)$
is the transition probability from state $j$ to state $j'$, where the
states are numbered in the order listed in \eqref{eq:list}. If
$f$ is chosen so that $f(a,\rho(a,k)) \in (0,1)$
for all $(a,k) \in \mathcal{S}$,
then the Markov chain is irreducible and there exists a unique
stationary distribution $\mathbf{v}(f)$ on
$\mathcal{S}$, represented by a row vector of length $2N$, that
satisfies
\begin{align*}
\mathbf{v}(f) = \mathbf{v}(f) \mathbf{Q}(f) \textrm{ and }
\mathbf{v}(f) \mathbf{e}_{2N} = 1,
\end{align*}
where $\mathbf{e}_{2N}$ is a column vector of length $2N$ whose
elements are all 1 \cite{meyer}. Let $s^* \triangleq (T,1)$ be the
state of a successful transmission. Then the throughput of user $i$ is given by the first
element of $\mathbf{v}(f)$, i.e.,
\begin{align*} 
\tau_i(f) = v(s^*;f).
\end{align*}
By symmetry, total throughput is given by $\tau(f) = N \tau_i(f)$.

Using \eqref{eq:avedel}, we obtain the relationship
\begin{align} \label{eq:expdeli}
\quad d_i(s;f) = 1 + \sum_{s' \in \mathcal{S} \setminus \{s^*\}} Q(s'|s;f)
d_i(s';f)
\end{align}
for all $s \in \mathcal{S}$.
Let $\mathbf{d}_i(f)$ be the column vector consisting of $d_i(s;f)$ and $\mathbf{Q}_0(f)$ be a matrix
obtained by replacing all the elements in
the first column of $\mathbf{Q}(f)$ by 0. Then \eqref{eq:expdeli} can be expressed as the following matrix equation:
\begin{align} \label{eq:matdel}
\mathbf{d}_i(f) = \mathbf{Q}_0(f) \mathbf{d}_i(f) +
\mathbf{e}_{2N}.
\end{align}
Solving \eqref{eq:matdel} for $\mathbf{d}_i(f)$, we obtain
\begin{align*}
\mathbf{d}_i(f) = (\mathbf{I}_{2N} - \mathbf{Q}_0(f))^{-1}
\mathbf{e}_{2N},
\end{align*}
assuming that the inverse exists, where $\mathbf{I}_{2N}$ is the $(2N \times 2N)$ identity matrix.
Since the long-run frequency of each state is given by the stationary
distribution $\mathbf{v}(f)$, the average delay of user $i$ can be expressed as
\begin{align*}
{D}(f) = \mathbf{v}(f) \mathbf{d}_i(f) - 0.5.
\end{align*}
Also, the average inter-packet time is given by the first element
of $\mathbf{d}_i(f)$, i.e., $\tilde{D}_i(f) = d_i(s^*;f)$.

\section{Derivation of Throughput and Average Delay
Under Symmetric Protocols With 1-Slot Memory and Memoryless Protocols in the
WLAN Model}

Let $f$ be a symmetric protocol with 1-slot memory.
The long-run fractions of idle, success, and
collision slots are given by
\begin{align}
P_0(f) &= v((W,0);f), \label{eq:p0dcf} \\
P_1(f) &= v((T,1);f) + v((W,1);f), \label{eq:p1dcf}\\
P_2(f) &= 1 - P_0(f) - P_1(f).\nonumber
\end{align}
Using \eqref{eq:dcf}, total throughput under $f$ in the WLAN model
can be written as
\begin{align} \label{eq:dcf22}
\tau(f) = \frac{P_{1}(f) E[P]}{P_0(f) \sigma_0 +
P_1(f) \sigma_1 + P_2(f) \sigma_2}.
\end{align}

In the WLAN model, we define the average delay of a user
as the average waiting time (measured in a time unit) until the beginning of
its next successful transmission starting from
an arbitrarily chosen time. We define $d_i(s;f)$ as the average
waiting time of user $i$ starting from the beginning of a slot
whose outcome yields state $s$ to user $i$.
Let $\sigma(s)$ be the duration of a slot yielding state $s$,
i.e., $\sigma(W,0) = \sigma_0$, $\sigma(a,1) =
\sigma_1$ for $a = T,W$, and $\sigma(a,k) = \sigma_2$ for $a = T,W$
and $k \geq 2$. Then (\ref{eq:expdeli}) can be modified as
\begin{align} \label{eq:expdeli2}
d_i(s;f) = \sigma(s) + \sum_{s' \in \mathcal{S} \setminus \{s^*\}} Q(s'|s;f)
d_i(s';f).
\end{align}
Let $\mathbf{b}$ be the column vector of the durations of slots, i.e.,
$\mathbf{b} \triangleq (\sigma(T,1) \ \sigma(T,2) \ \cdots \ \sigma
(W,N-1))^{\top}$. Then \eqref{eq:expdeli2} can be written as a
matrix equation
\begin{align} \label{eq:matdel2}
\mathbf{d}_i(f) = \mathbf{Q}_0(f) \mathbf{d}_i(f) +
\mathbf{b}.
\end{align}
Solving \eqref{eq:matdel2} for $\mathbf{d}_i(f)$, we obtain
\begin{align*}
\mathbf{d}_i(f) = (\mathbf{I}_{2N} - \mathbf{Q}_0(f))^{-1}
\mathbf{b}.
\end{align*}
The probability that an arbitrarily chosen time belongs to a slot yielding state $s$ is given by
\begin{align*}
y(s;f) = \frac{v(s;f) \sigma(s)}{\sum_{s' \in \mathcal{S}} v(s';f) \sigma(s')}.
\end{align*}
Note that a user stays in the initial slot for the half of its duration on
average.
Let $\mathbf{y}(f)$ be the row vector consisting of $y(s;f)$. Then
the average delay of user $i$ under protocol $f$ can be computed by
\begin{align*}
D(f) = \mathbf{y}(f) [ \mathbf{d}_i(f) - 0.5 \mathbf{b} ].
\end{align*}

Now let $f$ be a symmetric memoryless protocol, which is simply
a single transmission probability.
Then \eqref{eq:p0dcf} and \eqref{eq:p1dcf} can be expressed as
$P_0(f) = (1-f)^{N}$ and $P_1(f) = N f (1-f)^{N-1}$, respectively,
and we can use \eqref{eq:dcf22} to compute total throughput.
When users follow a memoryless protocol $f$,
the average waiting time starting from the next slot,
$d_i(s;f) - \sigma(s)$, is independent of $s$, and thus we
write the value as $\tilde{d}_i(f)$. Manipulating \eqref{eq:expdeli2}
yields
\begin{align*}
\tilde{d}_i(f) = \frac{N[P_0(f) \sigma_0 + P_2(f) \sigma_2] + (N-1)P_1(f) \sigma_1}{P_1(f)}.
\end{align*}
The average waiting time until the next slot is given by
\begin{align*}
r(f) = \frac{1}{2} \mathbf{y}(f) \mathbf{b} = \frac{P_0(f) \sigma_0^2 + P_1(f) \sigma_1^2 +
P_2(f) \sigma_2^2}{2 [P_0(f) \sigma_0 + P_1(f) \sigma_1 + P_2(f) \sigma_2]}.
\end{align*}
Hence, average delay under memoryless protocol $f$ can be computed by
\begin{align*}
D(f) = r(f) + \tilde{d}_i(f).
\end{align*}

\end{document}